\documentclass[twocolumn,twocolappendix]{aastex631}
\shorttitle{Eclipse mechanism of PSR J1544+4937}
\shortauthors{Kansabanik et al.}
\received{June 12, 2021}
\revised{July 27, 2021}
\accepted{July 27, 2021}
\published{October 13, 2021}
\begin{document}
\title{Unraveling the Eclipse Mechanism of a Binary Millisecond Pulsar Using Broadband
Radio Spectra}

\correspondingauthor{Devojyoti Kansabanik}
\email{dkansabanik@ncra.tifr.res.in}

\author[0000-0001-8801-9635]{Devojyoti Kansabanik}
\affiliation{National Centre for Radio Astrophysics, Tata Institute of Fundamental Research, Pune University, Pune 411007, India}

\author[0000-0002-6287-6900]{Bhaswati Bhattacharyya}
\affiliation{National Centre for Radio Astrophysics, Tata Institute of Fundamental Research, Pune University, Pune 411007, India}

\author[0000-0002-2892-8025]{Jayanta Roy}
\affiliation{National Centre for Radio Astrophysics, Tata Institute of Fundamental Research, Pune University, Pune 411007, India}

\author[0000-0001-9242-7041]{Benjamin Stappers}
\affiliation{Jodrell Bank Centre for Astrophysics, Department of Physics and Astronomy, The University of Manchester}

\begin{abstract}
The frequency dependent eclipses of the radio emission from millisecond pulsars (MSPs) in compact binary systems provide an opportunity to understand the eclipse mechanism and to determine the nature of the eclipsing medium. We combine multifrequency observations from the upgraded Giant Metrewave Radio Telescope (uGMRT) and model the broadband radio spectrum in the optically thick to thin transition regime to constrain the eclipse mechanism. The best-fit model to the eclipse phase spectra favors synchrotron absorption by relativistic electrons. We are able to strongly constrain the frequency of onset of the eclipse to 345$\pm$5 $\mathrm{MHz}$, which is an order of magnitude more precise than previous estimates. The dependence on the magnetic field strength of synchrotron absorption allowed us to estimate the magnetic field strength of the eclipse medium to be $\sim$13 $\mathrm{G}$, which is very similar to the values obtained by considering a pressure balance between the incident pulsar wind and the stellar wind of the companion. Applying this method to other millisecond binary pulsars will enable us to determine if the eclipse mechanisms are all the same and also estimate the wind and magnetic field properties of the companion stars. The method could also be applied to other systems where pulsars interact with companion winds in binary systems and in all cases it will lead to a better understanding of the evolutionary processes. 
\end{abstract}

\accepted{July 27, 2021}
\submitjournal{ApJ}

\section{Introduction}\label{sec:intro}
\begin{figure*}
    \centering
    \includegraphics[scale=0.9]{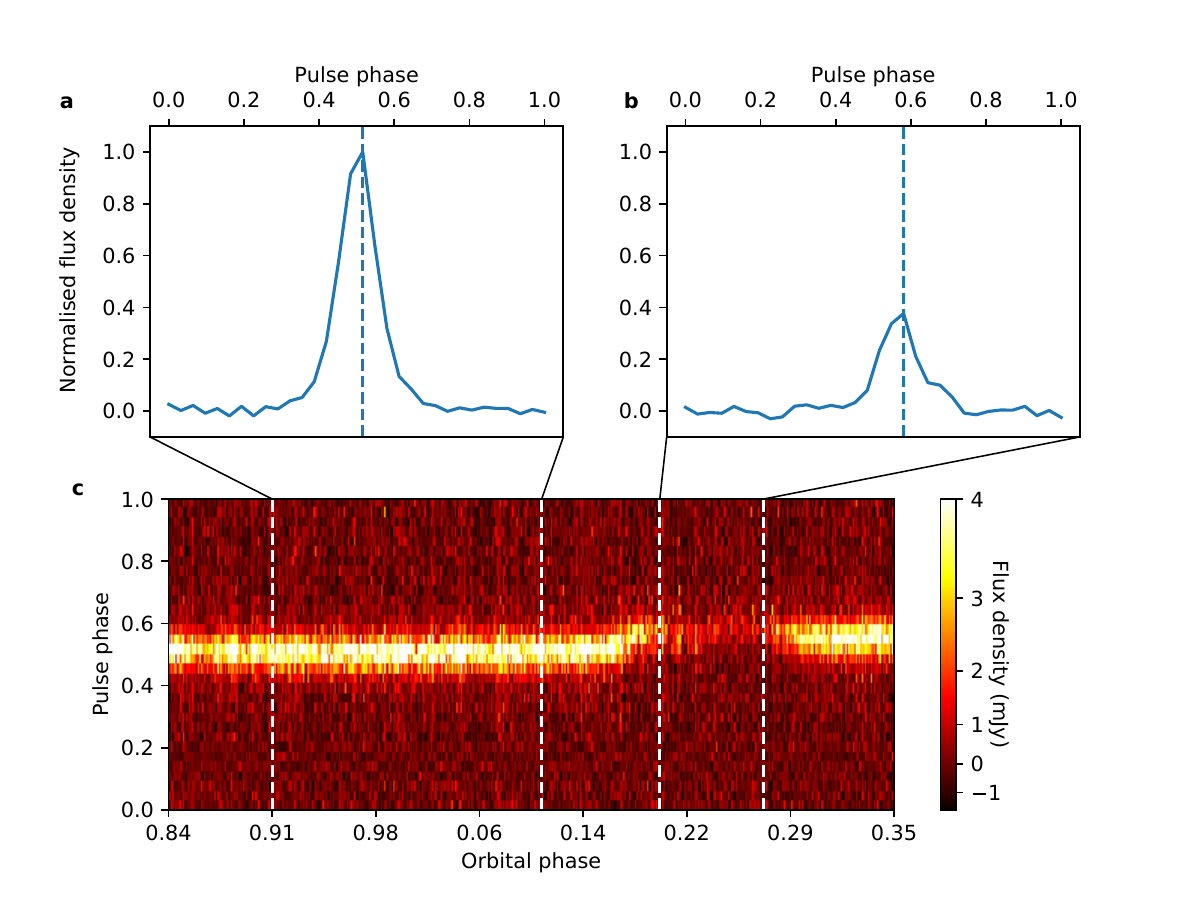}
    \caption{{\textbf{Pulsed flux density variation of PSR J1544$+$4937 in the FEP and NEP at band 3 on 2018 February 6. (a),(b)}} Average pulse profile in the NEP (orbital phase 0.91$-$0.10) and the FEP (orbital phase 0.19$-$0.27) are shown respectively. Both the pulse profiles in the NEP and FEP are normalized with respect to the NEP peak flux density. The flux density at the FEP is reduced about 60\% from the NEP flux density. The peak of the pulse profile is shifted $\sim100 \mathrm{\mu s}$ in the FEP (from pulse phase 0.5 to 0.55) due to the observed excess electron column density in the eclipse phase. {\textbf{(c)}} Variation of the pulsed flux density with orbital phase is shown. The flux density is shown by the color scale. At the FEP there is a significant reduction in the flux density associated with excess time delay in pulse arrival times.}
    \label{fig:fig1}
\end{figure*}

Millisecond pulsars (MSPs) are rapidly rotating neutron stars that are spun up to spin periods of a few milliseconds via the transfer of angular momentum through accretion of stellar material from the companion star \citep{Bhattacharya_1996}. MSPs found in compact binary orbits (orbital period; $P_b\leq 10$ $\mathrm{hours}$) around a low-mass companion ($M_C\leq 0.1 $ $M_\odot$) are important for understanding the formation of MSPs and the links between binary and isolated MSPs \citep{Roberts2013,Benvenuto_2014}. Most of the MSPs in compact orbits show frequency dependent eclipses of the radio emission from the pulsar when the companion star comes close to the line-of-sight (LOS). It is believed that eclipses are caused by either the material blown from the companion star by the pulsar wind or the material inside the pulsar wind itself. Following the discovery of the first eclipsing MSP B1957+20 \citep{Fruchter_1990}, eclipse mechanism studies were performed for a handful of eclipsing binary pulsars (e.g PSR J1227$-$4853 \citep{Roy_2015,Kudale_2020}, PSR J1544+4937 \citep{Bhattacharyya_2013}, PSR J1744$-$24A \citep{Lyne_1990}, PSR J1810+1744 \citep{Polzin_2018}, PSR B1957+20 \citep{Fruchter_1990,Fruchter1988,Polzin2020} and PSR J2051$-$0827 \citep{Stappers_1998,Polzin_2019}). The majority of the previous studies were limited to narrow bandwidth observations at the frequencies where the eclipse medium is optically thick and the pulsar radiation is completely eclipsed. Thus, these investigations could only probe the eclipse boundaries implying that the eclipse mechanism and the eclipse medium properties had to be inferred indirectly. Simultaneous multifrequency studies have also been undertaken, but they were typically at widely separated frequencies and the spectral evolution due to the eclipse medium was difficult to constrain. Detailed investigation of the frequency dependent eclipsing with wide bandwidth observations are therefore needed to probe the physical conditions near the superior conjunction \citep{Paulo_2004} which could be significantly different than near the eclipse boundaries. 

The majority of the previous studies concluded that the eclipse at lower frequencies ($\leq 1$ $\mathrm{GHz}$) may be caused by cyclotron/synchrotron absorption \citep{Thompson_1994}. Thus, the magnetic field is an important physical parameters of the eclipse medium in causing the eclipse by cyclotron/synchrotron absorption. Previous studies \citep{Fruchter_1990,Li2019,Crowter} estimated the magnetic field strength of the eclipse medium at the eclipse boundaries. The estimated magnetic field strength was a few milligauss and at least two orders of magnitude smaller than the characteristic magnetic field strength ($B_E$) calculated using the pressure balance between the pulsar wind and the stellar wind of the companion. One possible reason may be the different eclipse medium properties at different orbital phases. Electron column density variations with orbital phase in different eclipsing MSPs have been reported by previous studies \citep{Fruchter1988,Stappers_1998,Bhattacharyya_2013,Polzin_2018}. Similarly the magnetic field strength of the eclipse medium may also vary as we move away from the superior conjunction of the companion \citep{Khechinashvili2000}.
Since the previous estimations of magnetic field strengths were at the eclipse boundaries, it is necessary to estimate the properties of the eclipse medium directly at the superior conjunction.

The availability of the new wide bandwidth facilities like the Parkes ultra-wideband receiver \citep[UWL]{Hobbs_2020} and the upgraded Giant Metrewave Radio Telescope \citep[uGMRT]{Gupta_2017} allows one to probe the eclipse medium while transitioning from the optically thick to the optically thin regime. Although some previous studies \citep{Polzin_2019,Polzin2020} used wide bandwidth observations with the Parkes and uGMRT, they did not use the broadband radio spectrum to probe the eclipse mechanism. 
Here we demonstrate a new method that utilizes the broadband radio spectrum, a crucial discriminator between different eclipse mechanisms, for probing the eclipse mechanism.
We have applied this new method to the eclipsing MSP J1544$+$4937 \citep{Bhattacharyya_2013} to provide  strong constraints on the eclipse mechanism and also to estimate the physical parameters of the eclipse medium at the superior conjunction. The details of the observations and the data analysis are discussed in Section \ref{sec:obs}. In Section \ref{sec:result}, our findings on eclipse mechanisms and eclipse medium properties for PSR J1544$+$4937 are detailed. Section \ref{sec:dis} presents a summary and  discussion of the prospects for future implementation of this method to other systems.

\section{Observation and Data Analysis}\label{sec:obs}
A crucial discriminator between the different eclipse mechanisms is the frequency dependence of the eclipse duration, and it is apparent from the earlier narrow bandwidth observations that wide bandwidth observations are needed to capture any changes in the spectral properties of the received emission due to the eclipse medium. For this study we chose PSR J1544+4937 \citep{Bhattacharyya_2013}, an MSP discovered with the GMRT \citep{Swarup_1991}, which is in a compact binary orbit ($P_b\approx2.9$  $\mathrm{hours}$). Previous investigations with the GMRT software backend \citep{Roy2010} using 32 $\mathrm{MHz}$ of bandwidth indicates that PSR J1544+4937 exhibits frequency dependent eclipsing. Although the radio emission of the pulsar is obscured near the companion's superior conjunction at 306$-$338 $\mathrm{MHz}$, it is detected at 591$-$623  $\mathrm{MHz}$ \citep{Bhattacharyya_2013}. This previous study reported an uncertainty of 250 $\mathrm{MHz}$ on the frequency where radio emission from the pulsar is no longer detected. To more accurately determine this transition frequency we used the wide bandwidth capabilities of the uGMRT to observe PSR J1544+4937 at 300$-$500 $\mathrm{MHz}$ (band 3) and 650$-$850 $\mathrm{MHz}$ (band 4). We used high time resolution non-imaging data in total intensity for this study.

\subsection{Observation}\label{subsec:obs}
The observations were performed with the uGMRT \citep{Gupta_2017}, which is a radio interferometric array consisting of 30 dishes. We performed observation on three different epochs. Observations on 2018 February 6 and 2018 April 17 were performed by splitting the total number of antennas into two sub-arrays at 300$-$500 $\mathrm{MHz}$ and 650$-$850 $\mathrm{MHz}$, whereas the observations on 2018 May 7 were performed using all antennas at 300$-$500 $\mathrm{MHz}$. The observations were carried out in phased array mode where the spectral voltage signals from different antennas are coherently added together to form a single dish using the whole array. Coherent beam filterbank data at 48.28  $\mathrm{kHz}$ frequency resolution with 4096 frequency channels were recorded at every 81.92 $\mathrm{\mu s}$. The observations were scheduled in such a way that the full eclipse phase (FEP) was covered.

\subsection{Data processing}
We have used the GMRT pulsar tool [{\sc{gptool}} (Chowdhury A. et al., in preparation) to perform automated radio frequency interference (RFI) mitigation. The data were then corrected for interstellar dispersion with incoherent dedispersion \citep{Lorimer2004L} and then folded with the known ephemeris of the pulsar using {\sc{dspsr}} \citep{dspsr2010}. We have used {\sc{Tempo2}} \citep{Hobbs2006,Edwards2006} to calculate the difference between the observed and computed pulse time of arrivals (TOAs) using the known ephemeris of the pulsar derived from non-eclipse phase (NEP) TOAs. Then the excess electron column density in the eclipse medium is measured directly from the observed excess time delay of the pulsed emission  \citep{Lorimer2004L} as
\begin{equation}
    DM_{ex}(\mathrm{pc~cm^{-3}})= 2.4\times 10^{-10} t_{ex}(\mathrm{\mu s}) f(\mathrm{MHz})^2
\end{equation}
where $t_{ex}$ is the excess observed time delay, $DM_{ex}$ is the excess dispersion measure (DM) calculated from the $t_{ex}$ and $f$ is the observing frequency in  $\mathrm{MHz}$. The excess DM is then converted to excess electron column density as
\begin{equation}
    N_e(\mathrm{cm^{-2}}) = 3\times 10^{18}\times DM (\mathrm{pc~cm^{-3}})
\end{equation}
The observed data are flux calibrated using a flux calibrator 3C 286 (Appendix \ref{appendixa1}). We used the flux calibrated pulse averaged flux density for the rest of the work (Appendix \ref{appendixa2}).

\subsection{Variation of flux density and electron column density across the eclipse phase}\label{subsec:fluxvariation}
The flux density reduced during the eclipse phase. In Figure \ref{fig:fig1}c, we show the variation of pulsed flux density in band 3 as a function of orbital phase. We define the  orbital phase 0.19$-$0.27 around superior conjunction ($\phi_b$ = $0.25$) to be FEP and 0.91$-$0.10 to be a part of the NEP. There is an approximate 60\% reduction in the received pulse flux density in the FEP and a corresponding average delay in the arrival time of the pulses of 100 $\mathrm{\mu s}$ (Figure \ref{fig:fig1}a,b). In Figure \ref{fig:toa}a, we show the variation in the delay in the pulse TOA in band 3 and the corresponding excess electron column density throughout the eclipse phase for three different epochs (Section \ref{sec:obs}). The variation in flux density, normalized with respect to the NEP flux density, is shown in Figure \ref{fig:toa}b. We have marked three different regions in the eclipse phase; eclipse ingress phase (EIP), FEP and eclipse egress phase (EEP) in Figure \ref{fig:toa}. In contrast to what was seen in band 3, in band 4, we detect no change in the observed flux density between the FEP and NEP greater than our noise limit of $\sim100$ $\mathrm{\mu Jy}$.

\subsection{Spectra at different orbital phases}\label{subsec:spectramaking}
To further study the frequency dependence of the flux density of the pulsar during the eclipse phase, we have generated broadband spectra for PSR J1544+4937 at different orbital phases, namely, the NEP, FEP, EIP and EEP by splitting the observing band into smaller frequency slices. Different eclipse phases are shown in Figure \ref{fig:toa}. We divided the observing band into frequency slices of 10 $\mathrm{MHz}$ where the pulsar is still bright enough to potentially be detected in the NEP. In order to increase the signal-to-noise ratio during the brief ingress and egress phase we have averaged a 20 $\mathrm{MHz}$ bandwidth in both the EIP and EEP. The spectra at the NEP, FEP, EIP and EEP are shown in Figure \ref{fig:spectrum}.

\subsection{Spectrum Modeling}\label{subsec:spectramodel}
We modeled the FEP spectrum ($F_{ec}(\nu)$) for free-free absorption, induced Compton scattering, and synchrotron absorption. We used the radiative transfer equation $F_{ec}(\nu)=F_{nonec}(\nu)e^{-\tau(\nu)}$ to model the FEP spectrum, where $F_{nonec}$ is  the power-law spectrum fitted to the observed NEP spectrum; which is taken to be constant throughout the orbit. $\tau(\nu)$ is the frequency dependent optical depth for different eclipse mechanisms proposed by \cite{Thompson_1994}. We used the nonlinear curve fitting function {\it{curve\_fit}} of the {\it{SciPy}} package.

Analytical expressions of frequency dependent optical depths for the different mechanisms adopted from \cite{Thompson_1994} are used in this study.
\subsubsection{Free-Free Absorption}
The radio emission from the pulsar can be absorbed by the free electrons of the eclipse medium by the free-free absorption. The free-free absorption optical depth is given by 
\begin{equation}\label{eq:4}
    \tau_{ff}(\nu)\approx 3.8\times 10^{-14} \frac{f_{cl}}{T^{\frac{3}{2}}\nu^2L}N_{e}^2~ln(5\times 10^{10}\frac{T^{\frac{3}{2}}}{\nu})
\end{equation}
where, $T$ is the temperature of the eclipsing medium in $\mathrm{Kelvin}$, $N_{e}$ is the electron column density of the eclipse medium in $\mathrm{cm^{-2}}$, $f_{cl}=\frac{<n_e^2>}{<n_e>^2}$ is the clumping factor of the medium, and $L$ is the absorption length in centimeters. Assuming a spherical distribution of eclipse medium around the companion, we consider the maximum absorption length; $L_{max}=2R_E=1.0R_\odot$ as twice of the eclipse radius; $R_E=0.5R_\odot$. The observed FEP spectrum is fitted considering temperature, clumping factor, and absorption length as the free parameters using the expression for free-free optical depth in Equation \ref{eq:4}. During fitting we put a physically motivated constraint that the temperature $T>0$ and absorption length $L\leq L_{max}$.

\subsubsection{Induced Compton Scattering}
The induced Compton scattering optical depth is given by
\begin{equation}\label{eq:5}
    \tau_{ind}(\nu)\approx 4\times 10^{6} \frac{N_{e}S_{\nu}}{\nu^2}<f(\phi)>|\alpha+1|(\frac{d_{kpc}}{a})^2M 
\end{equation}
where $S_\nu$ is the mean flux density of the pulsar in millijansky at frequency $\nu$ $\mathrm{MHz}$, $\alpha$ is the spectral index of the incident radiation, $d_{kpc}$ is the distance to the scattering center from the observer in kiloparsecs and $a\sim1.2R_\odot$\citep{Bhattacharyya_2013} is the distance between the companion and the pulsar in centimeters. $f(\phi)$ is an angular factor which is averaged out over the scattering region \citep{Thompson_1994}. Reflection off a plasma cloud of radius of curvature $R_C$ will cause de-magnification $M\sim(\frac{R_C}{2r})^2$, where $r$ is the distance from the center of the curvature. The de-magnification factor varies between 0 and 1. We consider the NEP flux density as $6.5\pm0.2$ $\mathrm{mJy}$. The distance to the binary system estimated from optical observations is $2-5$ $\mathrm{kpc}$ \citep{Tang2014} and we take the average value; $d_{kpc}=3.5$ $\mathrm{kpc}$. The spectral index at the NEP is $\alpha_{nonec}$ = $-2.8\pm0.7$ and the distance between pulsar and companion is $a\sim1.2R_\odot$ \citep{Bhattacharyya_2013}. Considering these values and maximum de-magnification factor $M=1$, we have calculated the induced Compton optical depth.

\subsubsection{Cyclotron Absorption}
Since the companion has a magnetic field, cyclotron absorption is another possible eclipse mechanism. The cyclotron frequency is $\nu_B=\frac{eB}{2\pi m_ec}$ and the corresponding cyclotron harmonic at frequency $\nu$ is $m=\frac{\nu}{\nu_B}$. There are two components of cyclotron optical depths for two polarization components of the incident radiation. One component optical depth ($\tau_{\parallel}$) is parallel to the $(\vec{k}\times\vec{B})\times \vec{k}$ and another component ($\tau_{\perp}$) is perpendicular to the $(\vec{k}\times\vec{B})\times \vec{k}$, where $\vec{k}$ is the wavevector, $\vec{B}$ is the magnetic field vector. Considering the LOS angle with the magnetic field is 90 $\mathrm{degrees}$, $\tau_{\parallel}$ is very small compared to the $\tau_{\perp}$ at higher cyclotron harmonics. Thus, considering the LOS angle with the magnetic field is 90 $\mathrm{degrees}$, the cyclotron absorption optical depth for perpendicular polarization component corresponding to cyclotron harmonic $m$ is given by \citep{Thompson_1994} as
\begin{equation}\label{eq:6}
    {\tau_{abs}^{(m)}}_{\perp,\frac{\pi}{2}}=\frac{\pi}{2}\frac{m^{m+1}}{m!}(\frac{mk_BT}{2m_ec^2})^{(m-1)}\frac{n_ee^2L_B}{m_ec\nu}
\end{equation}
where and $L_B=|\frac{ds}{dlnB}|$ is called the scale length of magnetic variation. For simplicity, we consider the scale length of electron density variations and magnetic field variations as similar. Thus, the quantity $n_eL_B$ is taken to be equal to the average electron column density $N_e$ \citep{Thompson_1994}. The cyclotron approximation is valid if the temperature $T\leq\frac{m_ec^2}{2km^3}$ \citep{Thompson_1994}. 

\subsubsection{Synchrotron Absorption}
In the trans-relativistic case, thermal electrons dominate the absorption if cyclotron harmonic is $m<\frac{1}{2}(p+1)$ \citep{Thompson_1994}. Thus, thermal electrons dominate the absorption at lower cyclotron harmonics. At higher harmonics, we consider only the synchrotron absorption by nonthermal electrons with energy density distribution $n(E)=n_0E^{-p}; E_{min}<E<E_{max}$. The optical depth is given by 
\begin{equation}\label{eq:7}
    \tau_{syn}=(\frac{3^{\frac{(p+1)}{2}}\Gamma(\frac{3p+2}{12})\Gamma(\frac{3p+22}{12})}{4})(\frac{sin\theta}{m})^{\frac{p+2}{2}}\frac{n_0e^2}{m_ec\nu}L
\end{equation}
where $L$ is the absorption length and $p$ is the power-law index. Here we have considered an average magnetic field strength along the LOS and homogeneous distribution nonthermal electron. We considered a tiny fraction ($\sim$1\%) of total electron density ($n_e$) as a typical value of nonthermal electron density \citep{Thompson_1994}. For a chosen value of LOS angle ($\theta$) we fitted the observed spectrum with $B$ and $p$ as free parameters.

\section{Results}\label{sec:result}
In this section we explore the eclipse mechanism and eclipse medium properties using the fitted broadband spectrum of the PSR J1544+4937 at different orbital phases around the superior conjunction.
\begin{figure*}
    \centering
    \includegraphics[trim={1cm 1cm 0cm 1cm},clip,scale=0.82]{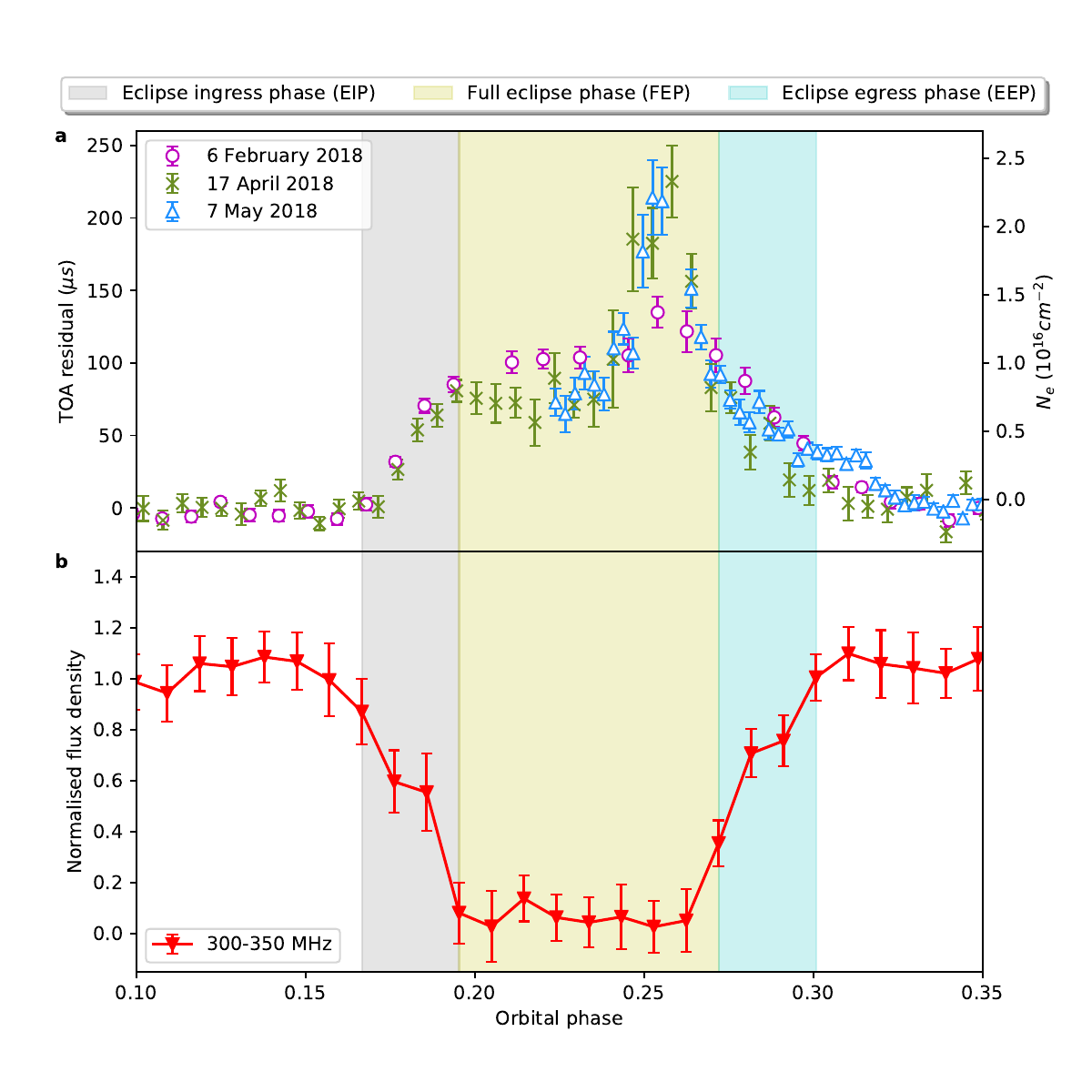}
    \caption{{\textbf{Variation of the excess electron column density ($N_e$) and the mean flux density with orbital phase around the superior conjunction.}} Considering the 300$-$350 $\mathrm{MHz}$ part of the band 3, the mean flux density drops to 85\% of the NEP flux density at orbital phase 0.16 and continues to drop to the 5$\sigma$ detection limit (1.1 $\mathrm{mJy}$) at orbital phase 0.19. The orbital phase range 0.16$-$0.19 shaded by gray denotes the EIP. The flux density again increased above 5$\sigma$ limit at orbital phase 0.27. The orbital phase range 0.19$-$0.27 shaded by yellow enotes the FEP. The flux density increased more than 85\% of the NEP flux density at orbital phase 0.30. The orbital phase range 0.27$-$0.30 shaded by sky blue denotes the EEP. {\textbf{(a)}} The observed delay at the center frequency of band 3 (400 $\mathrm{MHz}$) for the three different epochs is shown. The excess electron column density in the eclipse phase causes the observed delay in the pulse arrival time. {\textbf{(b)}} The mean flux density variation over 300$-$350 $\mathrm{MHz}$ normalized with respect to the NEP flux density is shown by red lower triangles.}
    \label{fig:toa}
\end{figure*}

\begin{figure*}
\centering
    \includegraphics[trim={0cm 1cm 0cm 1cm},clip,scale=0.86]{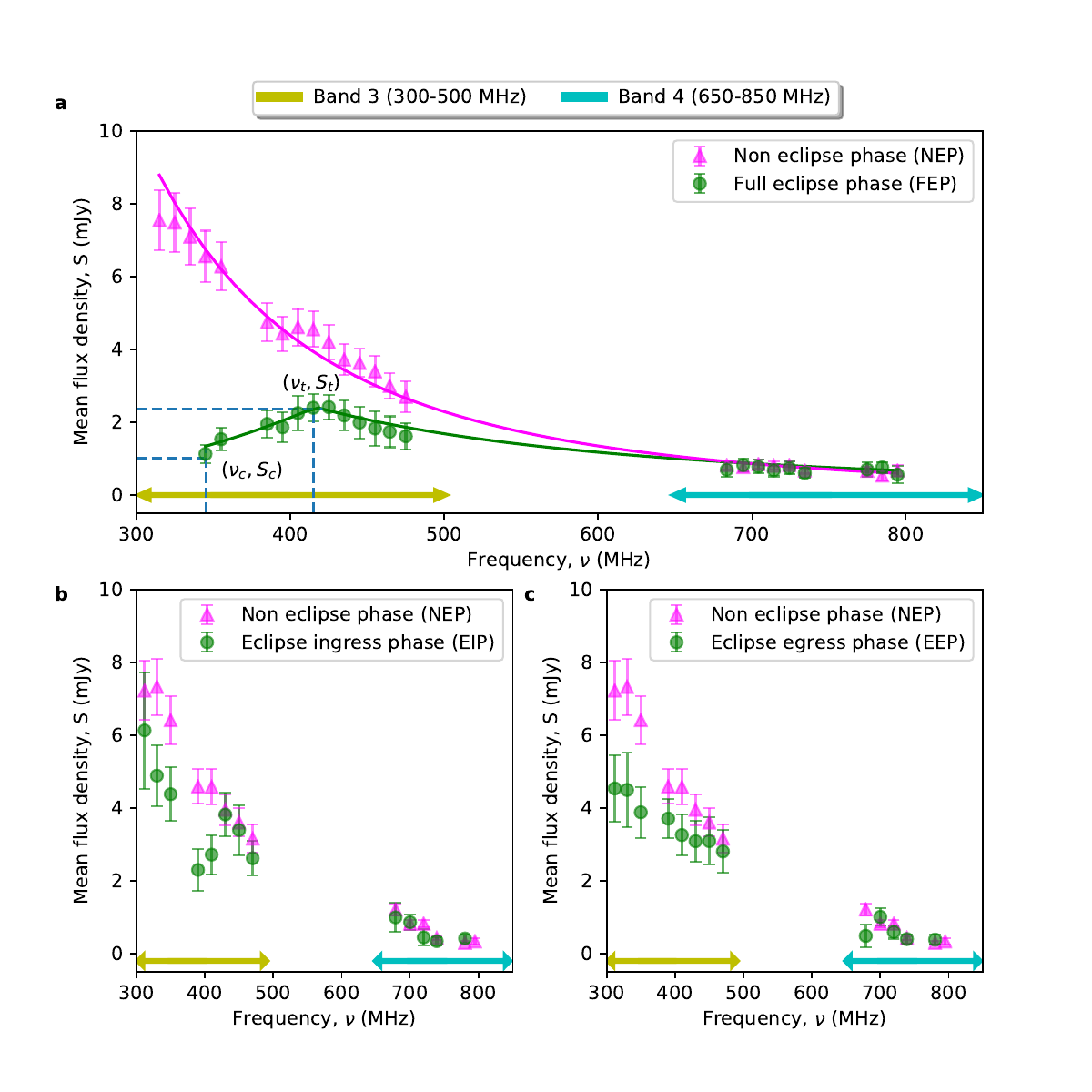}
    \caption{{\textbf {Spectral properties of the pulsed radio emission of the PSR J1544+4937 at the NEP, FEP, EIP, and EEP for the 2018 February 6 observation.}} The flux density as a function of frequency is shown. In all the plots the magenta points correspond to the mean flux densities measured at the NEP. The green points correspond to the mean flux densities measured in three eclipse regions; FEP (a), EIP (b), EEP (c). The span of the band 3 and band 4 are shown by the double headed arrows. Certain parts of the bands are masked due to the RFI or low sensitivity. {\textbf{(a)}} Fitted power law to the observed NEP spectrum is shown by the magenta solid line and the broken power law fitted to the observed FEP spectrum is shown by the green solid line. In the FEP the pulsar is not detected with a detection significance more than 5$\sigma$ below $\nu_c = 345\pm5$\ $\mathrm{MHz}$. The flux density at $\nu_c$ is  $S_c =1.1\pm0.2$\,$\mathrm{mJy}$. In the FEP spectrum we can also see that there is a turnover at $\nu_t = 416\pm33$\,$\mathrm{MHz}$ and the corresponding mean flux density is $S_t =2.4\pm0.3$\,$\mathrm{mJy}$. {\textbf{(b),(c)}} The spectra at the EIP and EEP are different in characteristics compared to the spectrum at the FEP. In order to increase the signal-to-noise ratio during the brief ingress and egress phase, we averaged a 20 $\mathrm{MHz}$ bandwidth in both the EIP and EEP.}
    \label{fig:spectrum}
    \end{figure*}

\subsection{Electron density at the Ingress/Egress Phase} 
The flux density drops to 85\% of the NEP phase flux density at orbital phase $\phi_i$ = 0.16 and $\phi_e$ = 0.30, where $\phi_i$ and $\phi_e$ are defined as the start and end of the EIP and EEP respectively (Figure \ref{fig:toa}b). The orbital phase range $\phi_e-\phi_i$ corresponds to the eclipse radius $R_E\sim0.5$ $R_\odot$. We observed the excess electron column density at the ingress and egress as $N_e\sim1.0\times10^{16}$ $\mathrm{cm^{-2}}$ (Figure \ref{fig:toa}a). Considering the constant electron density along the LOS, we estimate an average electron density; $n_e\approx1.5\times10^5$ $\mathrm{cm^{-3}}$ at the eclipse boundaries \citep{Polzin_2018}.

\subsection{Frequency Onset of the Eclipse}
Using the 10 $\mathrm{MHz}$ frequency slices, we find that in the FEP the pulsar is not detected above 5$\sigma$ at frequencies below an eclipse onset frequency of $\nu_c$ = 345$\pm$5 $\mathrm{MHz}$ where the pulsar has a flux density of $S_c=1.1\pm0.2$ $\mathrm{mJy}$ (Figure \ref{fig:spectrum}a). The error on the eclipse onset frequency is defined as the half width of the frequency slice. The estimated eclipse onset is an order of magnitude more precise $\nu_c$ compared to previous estimates for any eclipsing MSP \citep{Bhattacharyya_2013,Polzin_2019}.

\subsection{Observed Optical Depth at Eclipse Onset Frequency}
Due to the large uncertainty of $\nu_c$, previous studies could only put a limit on the physical parameters of the eclipse medium. With an order of magnitude of a more precise $\nu_c$ we can calculate the observed optical depth at the $\nu_c$. We use an average value of the observed electron column density in the FEP of $N_e=1.5\times10^{16}$ $\mathrm{cm^{-2}}$ (Figure \ref{fig:toa}a). We calculate the optical depth $\tau_{\nu,c}\approx1.7$ at $\nu_c$ since the flux density of the pulsar decreases from $6.5\pm0.2$ $\mathrm{mJy}$ in the NEP to $1.1\pm0.2$ $\mathrm{mJy}$ in the FEP (Figure \ref{fig:spectrum}a). This direct estimation of the optical depth at the FEP allows us to provide a strong constraint on the physical parameters of the eclipse medium, which was not possible before.

\subsection{Broadband Spectra at Different Orbital Phases}
At $\nu_c$ the eclipse medium is transitioning from an optically thick to thin regime which is hitherto unexplored for any other eclipsing MSP systems, since the majority of previous studies were performed at frequencies corresponding to the optically thick regime. In Figure \ref{fig:spectrum}a, we show the spectrum at the FEP and NEP for the observation on 2018 February 6. Fitting a power-law, $S_{nonec}(\nu)\propto\nu^{\alpha_{nonec}}$, where  $\alpha_{nonec}$ = $-2.8\pm0.7$ is the spectral index of the NEP spectrum. By comparison, the spectrum in the FEP has a break in the spectrum. We fit a broken power law to the observed FEP spectrum given as
\begin{equation}
    S(\nu) = \Bigg\{
        \begin{array}{ll}
        S_t(\frac{\nu}{\nu_t})^{\alpha_{low}}, & \nu<\nu_t,\\
        S_t(\frac{\nu}{\nu_t})^{\alpha_{high}}, & \nu>\nu_t
        \end{array}
\end{equation}
where $\nu_t$ is the turnover frequency, $S_{t}$ is the peak flux density at the turnover frequency $\nu_t$, $\alpha_{low}$ is the spectral index for frequencies below $\nu_t$ and $\alpha_{high}$ is the spectral index for frequencies above $\nu_t$. The fitted power law spectrum for the NEP is shown by magenta solid line and the fitted broken power law for the FEP is shown by the green solid line in Figure \ref{fig:spectrum}a. The errors on $\nu_t, \alpha_{low}, \alpha_{high}$ and $S_t$ are the errors from fitting. Above a turn over frequency of $\nu_t$ = $416\pm33$ $\mathrm{MHz}$ the spectral index $\alpha_{high}=-1.9\pm0.2$ and below $\nu_t$ the spectrum has a positive spectral index, $\alpha_{low}=3.1\pm0.1$. The peak flux density at $\nu_t$ obtained from broken power-law fit is $S_t=2.4\pm0.3$ $\mathrm{mJy}$ This difference between the NEP and the FEP spectral shape has not been reported before in any other eclipsing binary MSP. A comparison between the NEP spectrum with those in the EIP and the EEP are shown in the lower panels of Figure \ref{fig:spectrum}, and we can see that there are differences compared to the FEP spectrum suggesting that there may be different eclipse mechanisms in place in the different regions, or the physical properties of the material are sufficiently different. 

\begin{figure*}
    \centering
    \includegraphics[trim={1.6cm 0.0cm 2.0cm 0.8cm},clip,scale=0.85]{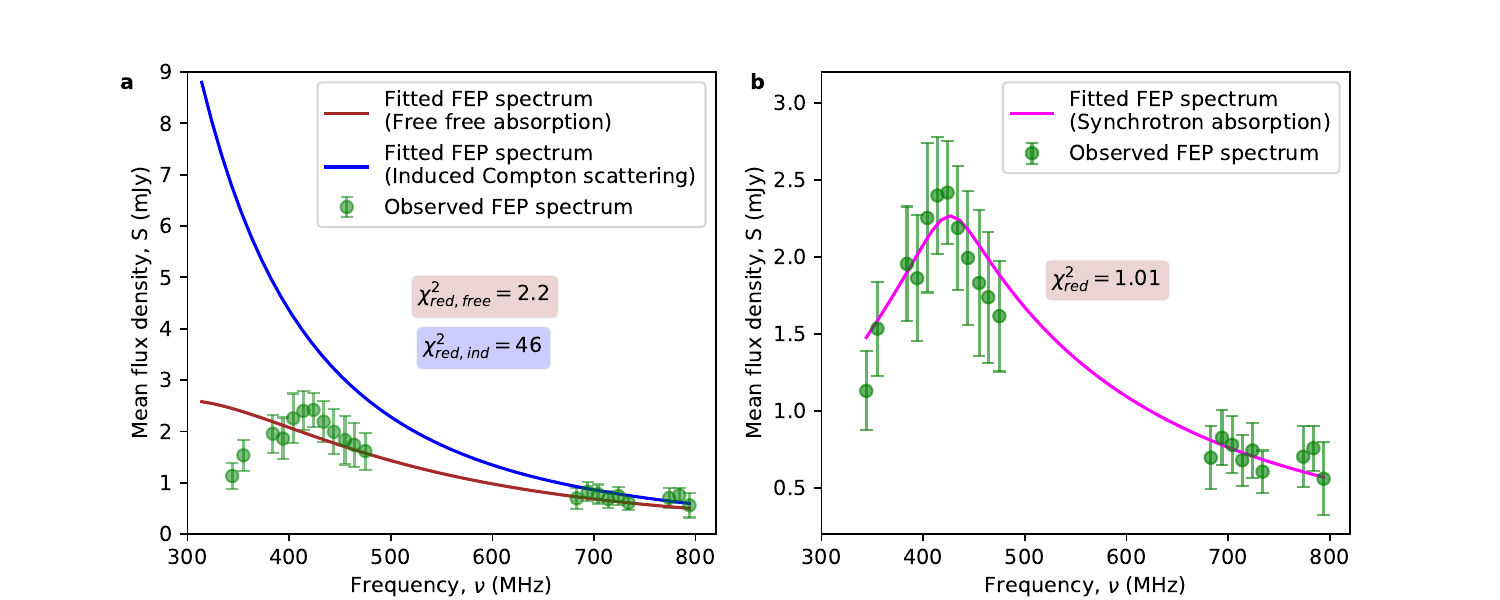}
    \caption{{\textbf{Observed FEP spectrum on 2018 February 6 fitted with the free-free absorption, the induced Compton scattering, and synchrotron absorption models.}} The observed FEP spectrum is shown by green circles in both the figures. {\textbf{(a)}} Fitted spectrum is shown by the brown solid line for the free-free absorption model and by the blue solid line for the induced Compton scattering. {\textbf{(b)}} The best-fit spectrum considering the synchrotron absorption by the relativistic electrons is shown by the magenta solid line. The observed FEP spectrum is well fitted with this model (the reduced chi-square is 1.01).}
    \label{fig:modelspectra}
\end{figure*}
\begin{figure}
      \centering
      \includegraphics[trim={0cm 0cm 0cm 1cm},clip,scale=0.56]{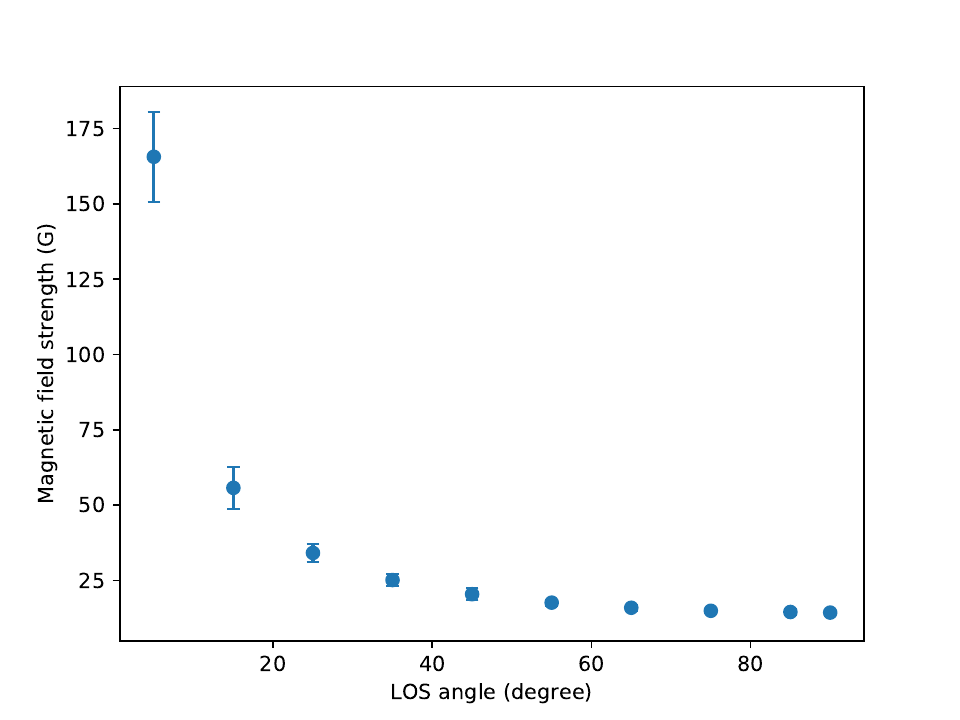}
      \caption{\textbf{Estimated magnetic field strength considering different LOS angles.} The average magnetic field strength estimated (B) during the fitting of the synchrotron model with the FEP spectrum has a degeneracy with the LOS angle ($\theta$). While fitting the FEP spectrum for the 2018 February 6 observation, we vary the LOS angle between 5$^\circ$ and 90$^\circ$ in 10$^\circ$ step. We have kept the LOS angle fixed during each fitting and the corresponding estimated magnetic field strengths are shown in this figure.}
    \label{fig:LOS_B}
  \end{figure}
\subsection{Eclipse Mechanism}\label{sec:ecmech}
We consider different possible eclipse mechanisms \citep{Thompson_1994} to explain our observations (Section \ref{subsec:spectramodel}). Radiation at a frequency below the plasma frequency, $\nu_p$ = $8.5$ $(n_e/\mathrm{cm^{-3}})$ $\mathrm{kHz}$, of the eclipsing material will not be able to propagate through it. If we consider that the observed $\nu_c$ corresponds to $\nu_p$ in the FEP we find that the corresponding electron density is $n_{e,p}=1.6\times10^{9}$ $\mathrm{cm^{-3}}$. This is in stark contrast to the electron density in the eclipse region, $n_e\sim1.5\times10^5$ $\mathrm{cm^{-3}}$,  estimated by assuming a spherically distributed eclipse medium and the electron densities derived from the delayed pulse arrival times (Figure \ref{fig:toa}a). This suggests that the eclipse onset at $\nu_c$ cannot be explained by the plasma frequency cutoff. 
On the other hand, radiation at frequencies higher than $\nu_p$ undergo refraction due to the intervening medium. If the angle of refraction is sufficiently large, the pulsed emission will be refracted out of the LOS resulting in an eclipse. If refraction is the cause of the observed eclipse at $\nu_c$, the delay of the pulse arrival time at the eclipse boundaries is $\sim$10 $-$ 100 ms \citep{Thompson_1994}, which is much larger than the observed delay in pulse arrival time at the eclipse boundaries $\sim$ $100 \mathrm{\mu s}$ (Figure \ref{fig:toa}a). This suggests that the refraction at the frequencies below $\nu_c$ is not sufficient to explain the observed eclipse. Pulse broadening due to excess DM and scattering in the FEP is $\sim45$ $\mathrm{\mu s}$ at $\nu_c$ (Appendix \ref{subsec:appendixscatter}), which is less than the time resolution of the data, $81.92$ $\mathrm{\mu s}$ and so it cannot reduce the observed pulse detection.

\subsection{Modeled Spectrum}
We have then modeled the spectrum in the FEP using the radiative transfer equation (Section \ref{subsec:spectramodel}) for three mechanisms: free-free absorption, induced Compton scattering and synchrotron absorption. We have assumed that the spectra at the NEP does not change for other reasons. We consider a given mechanism as the most plausible when the observed optical depth ($\tau_{\nu,c}$) and the observed FEP spectrum can be reproduced with physically acceptable parameters. We have modeled the FEP spectra for three different epochs to see whether eclipse mechanism or the eclipse medium properties varies significantly between epochs.

\subsubsection{Free-Free absorption}\label{subsec:free}
The free-free absorption optical depth ($\tau_{ff}$) [Equation \ref{eq:4}] due to the free electrons in the eclipse medium  depends on three physical quantities: temperature ($T$), absorption length ($L$) and clumping factor ($f_{cl}=\frac{<n_e^2>}{<n_e>^2}$). Clumping factor is a measure of inhomogeneity of the medium. To obtain the observed value of $\tau_{\nu,c}$ from free-free absorption the temperature of the medium would need to be $T=130$ $(f_{cl})^{\frac{2}{3}}$. Thus, either a very low temperature ($T\sim$1000 $\mathrm{K}$) or high clumping factor ($f_{cl}\sim10^9$) is required to create an eclipse at $\nu_c$ due to free-free absorption. We fit (brown solid line in Figure \ref{fig:modelspectra}a) the observed spectrum in the FEP (green circles in Figure \ref{fig:modelspectra}a) using the frequency dependent optical depths of free-free absorption but did not obtain a good fit ($\chi^2_{red}\sim2$). The mismatch between the observed and modeled FEP spectrum suggests that the free-free absorption cannot produce the observed frequency dependent eclipses of PSR J1544+4937. 

\subsubsection{Induced Compton Scattering}\label{subsec:comp}
Another possible eclipse mechanism is induced Compton scattering, which is a nonlinear scattering effect. Although the total number of incident photon remain conserved, this nonlinear scattering effectively changes the spectrum of the incident radiation and induces an optical depth ($\tau_{ind}$) [Equation \ref{eq:5}]. We have calculated $\tau_{ind}=4\times 10^{-2}$ at $\nu_c$, which is two orders of magnitude smaller than the observed $\tau_{\nu,c}$. We have also modeled the observed spectrum in the FEP with induced Compton scattering considering the de-magnification factor can vary between 0 and 1. We have found that the model spectrum (blue solid line in Figure \ref{fig:modelspectra}a) differs significantly to the observed FEP spectrum (green circles in Figure \ref{fig:modelspectra}a). The corresponding reduced $\chi^2$ is also very high; i.e. $\chi^2_{red}\sim46$.

\subsubsection{Possibility of Cyclotron Absorption}\label{subsec:cyclo}
Since the companion star and the pulsar wind have magnetic fields, we consider cyclotron absorption by nonrelativistic electrons as one of the possible eclipse mechanisms. Assuming an isotropic pulsar wind, we have estimated the characteristic magnetic field strength equating the magnetic pressure ($\frac{B_E^2}{8\pi}$) of the plasma of the eclipse medium to stagnation pressure of the incident pulsar wind, $U_E=\frac{\dot{E}}{4\pi ca^2}$, where $\dot{E}$ = $1.15\times 10^{34}$ $\mathrm{erg/s}$ is the spin down power of the pulsar and $a\sim1.2$ $R_\odot$ is the distance between pulsar and companion. We have calculated the pulsar wind energy density; $U_E=4.37$ $\mathrm{erg/cm^3}$ and the characteristic magnetic field strength is $B_E\approx 10$ $\mathrm{G}$. The corresponding cyclotron frequency is $\nu_B\approx 30$ $\mathrm{MHz}$ and the cyclotron harmonic corresponding to $\nu_c$ is $m\approx12$. This means that cyclotron absorption can only produce the observed eclipse at $\nu_c$ if the temperature is $T=7.9\times10^7$ $\mathrm{K}$ [Equation \ref{eq:6}]. However the cyclotron approximation is only valid for temperature $T\leq1.7\times10^{6}$ $\mathrm{K}$ \citep{Thompson_1994}. Thus, the required temperature is outside the range of cyclotron limit. Therefore, we rule out the cyclotron absorption by nonrelativistic electrons as the eclipse mechanism.

\subsubsection{Synchrotron Absorption}\label{sub:synchro}
In the trans-relativistic regime both the thermal and nonthermal electrons can contribute to the synchrotron absorption. Thermal electrons will dominate the absorption for lower cyclotron harmonic number $m<4$ \citep{Thompson_1994} considering a nonthermal electron power-law index $p=7$. But at higher cyclotron harmonic corresponding to $\nu_c$ ($m\approx12$) the absorption by the thermal electrons is negligible. Thus, we have fitted the observed FEP spectrum considering only the synchrotron absorption by relativistic nonthermal electrons with the energy distribution $n(E)=n_0E^{-p}$; $E_{min}<E<E_{max}$. We chose magnetic field strength $B=1$ $\mathrm{G}$ and $p=5$ as the initial guess values for performing the fitting. The modeled spectrum with synchrotron absorption (magenta solid line in Figure \ref{fig:modelspectra}b) matches the observed spectrum in the FEP very well and the reduced $\chi^2$ is also close to unity; $\chi_{red}^2=1.01$. 

The FEP spectrum is therefore best modeled with synchrotron absorption by relativistic electrons as the eclipse mechanism for the observing epoch 2018 February 6. We have also performed spectral modeling for PSR J1544+4937 on two additional epochs, 2018 April 17 and 2018 May 7, and found that the best fits are also achieved by synchrotron absorption. 
 
\subsection{Estimated Physical Parameters of Eclipse Medium}
While modeling the observed FEP spectrum with synchrotron absorption discussed in Section \ref{sub:synchro}, we assumed that the nonthermal electron distribution in the eclipse medium is homogeneous and the average LOS angle to the magnetic field is $90^\circ$. Best-fit values of $B$ and $p$ for the three epochs are given in Table \ref{table:table1}, which are broadly consistent. We obtain a mean value of magnetic field strength,  B $\sim13.7\pm1.2$ $\mathrm{G}$ across the three epochs, which is similar to the characteristic magnetic field strength $B_E$. We have estimated the nonthermal electron energy density, $U_{nth}\approx10^{-4}$ $\mathrm{erg/cm^3}$, which is only a tiny fraction of the total pulsar wind energy density, $U_E=4.37$ $\mathrm{erg/cm^3}$. The estimated LOS averaged magnetic field strength using the synchrotron absorption model depends on the LOS angle ($\theta$) with the magnetic field (Equation \ref{eq:7}). To explore further we fit the FEP spectra for different values of $\theta$ between the range of $5^\circ-90^\circ$ in $10^\circ$ steps and the estimated values of $B$ are shown in Figure \ref{fig:LOS_B}. The upper and lower limits of the estimated LOS averaged magnetic field strengths are $162\pm14$ $\mathrm{G}$ and $14.0\pm1.1$ $\mathrm{G}$ respectively, which are also a similar order of magnitude to $B_E$.

\begin{table}
\begin{tabular}{|c|c|c|}
    \hline
         Date of obs & $B(\mathrm{G})$ & $p$ \\[0.1cm]
         \hline
         2018 February 6 & $14.0\pm1.1$ & $4.7\pm0.6$ \\[0.1cm]
         \hline
         2018 April 17 & $17.3\pm1.1$ & $5.6\pm1.1$ \\[0.1cm]
         \hline
         2018 May 7 & $9.9\pm1.6$ & $3.7\pm0.8$ \\[0.1cm]
         \hline
    \end{tabular}
    \caption{\textbf{Note : Estimated physical parameters of the eclipse medium.} Two physical parameters; the LOS averaged magnetic field strength ($B$) and $p$  estimated during the fitting of the observed FEP spectrum are listed in this table. We consider the synchrotron absorption by relativistic non-thermal electrons following energy distribution $n(E)=n_0E^{-p}$ as the eclipse mechanism. We considered the magnetic field is perpendicular to the LOS during the estimation of these parameters. The estimated parameters for three different epochs are broadly consistent.}
    \label{table:table1}
\centering
\end{table}

\section{Discussion}\label{sec:dis}
This paper reports the very first modeling of the broadband radio spectra at the FEP to constrain the eclipse mechanism. We have found that the observed frequency dependent eclipses for PSR J1544+4937 can be well explained by synchrotron absorption by relativistic electrons, while the other mechanisms could not explain the observed eclipse (Section \ref{sec:ecmech}). We ruled out cyclotron absorption because the required temperature is much higher than the cyclotron limit for the observed optical depth at the $\nu_c$. We find that the synchrotron absorption can reproduce the observed FEP spectra and also the estimated physical parameters are feasible. Thus, we conclude that the synchrotron absorption is the most plausible eclipse mechanism for PSR J1544+4937.

We have shown that the observed spectra at three different orbital phases (EIP, FEP and EEP) are significantly different (Section \ref{subsec:spectramaking}). This implies that the frequency dependence of the optical depth is varying as a function of orbital phase. Different frequency dependence of the optical depth for the EIP, FEP and EEP indicates that either eclipse mechanisms or the eclipse medium properties are varying with the orbital phase. Sensitive orbital phase resolved studies are needed to probe this in detail.

\subsection{Fractional Contribution from Other Mechanisms}
When considering the influence of the different eclipse mechanisms we assumed that the spectrum in the NEP is representative of that seen throughout the orbit, if there were no eclipse. Scintillation could change the flux density of the pulsar; however, the decorrelation bandwidth for diffractive scintillation is $\sim1.1$ $\mathrm{kHz}$ (Section \ref{subsec:appendixscatter}) given the DM of the pulsar and thus will be averaged out over the 10 $\mathrm{MHz}$ frequency slices. Moreover, we would not expect scintillation to have this sort of orbital phase dependence. Although we have ruled out free-free absorption and induced Compton scattering as the main eclipse mechanism, we now consider if they may make a significant fractional contribution to the eclipses. Considering the average electron column density $N_e\sim$ $1.5\times10^{16}$ $\mathrm{cm^{-2}}$ at the FEP, and a typical temperature of the stellar wind \citep{Thompson_1994}; $T\sim10^8-10^9$ $\mathrm{K}$ and clumping factor; $f_{cl}\sim1$, we have estimated the free-free optical depth; $\tau_{ff}\sim10^{-9}-10^{-10}$ and the induced Compton scattering optical depth is $\sim4\times10^{-2}$ at the $\nu_c$. Thus, both free-free absorption and induced Compton scattering do not contribute significantly to the observed optical depth. Synchrotron absorption is therefore solely responsible for the observed frequency dependent eclipsing and the estimated physical parameters are not affected by a contribution to the optical depth by other mechanisms.

\subsection{Comparison of Estimated Magnetic Field Strength with Previous Studies}
Most previous studies \citep{Fruchter_1990,Polzin_2018,Fruchter1988,Polzin2020,Stappers_1998,Polzin_2019} have concluded that the low frequency eclipses may be caused by cyclotron/synchrotron absorption. The magnetic field strength, $B$, is a crucial parameter for cyclotron/synchrotron absorption. Recent studies of PSR B1957+20 \citep{Li2019} and PSR J2256$-$1024 \citep{Crowter} estimated $B$ of a few milligauss at the eclipse boundaries, which is significantly different from $B_E$ and insufficient to cause cyclotron absorption at the observing frequency. However, recent polarization observations of PSR J2051$-$0827 \citep{Polzin_2019} using the Parkes UWL \citep{Hobbs_2020} at the FEP provides a limit on the parallel component $B_\parallel$ = $20\pm120$ $\mathrm{G}$ of the magnetic field, which is close to the characteristic magnetic field strength for this pulsar. Our estimate of the magnetic field strength, using a completely different method for PSR J1544+4937, is consistent with $B_E$. Thus, the measurement of a very low magnetic field strength at the eclipse boundaries were possibly due to the observations at the eclipse boundaries where the eclipse medium properties may be significantly different than in the FEP \citep{Khechinashvili2000}. 

\subsection{Future Applications on Broad Range of Binaries}\label{subsec:future}
Our new method can be utilized to understand the eclipse mechanisms and the evolution of other eclipsing MSPs in compact binaries, including the transitional MSP J1227$-$4853 \citep{Roy_2015,Kudale_2020}. In the future more sensitive observations will allow us to perform the spectral modeling with higher orbital phase resolution covering the eclipse region. This method can also be used for a diverse range of binaries, with different companion types and orbital properties, for which spectral modulation and DM variation with orbital phase is, or expected to be, observed. For example, this method can probe the eclipse properties of PSR B1259$-$63 \citep{Johnston2004}, which is orbiting a massive Be star and the observed eclipses are thought to be caused by the circumstellar disk of the Be star and related systems, such as PSR J1740$-$3052 \citep{Madsen2012}. For the double pulsar binary, PSR J0737$-$3039, where the observed flux density of one pulsar is modulated by the second pulsar \citep{Lyne2004,Mc2004,Breton_2012}, an adoption of our simultaneous broadband observing approach may be used to probe the magnetosphere of the second one. Thus, even though primarily developed for the study of the eclipse properties of the MSPs in compact orbits, this technique has the potential to probe the stellar environment for a wide range of binaries via the modeling of broadband radio spectra. 

\appendix
\section{Appendix}\label{appendixa}
The method for flux calibration and estimation of mean flux densities are described in Appendix \ref{appendixa1} and Appendix \ref{appendixa2} respectively. We describe the calculation of the effect of pulse broadening and scintillation on the observed pulse width in Appendix \ref{subsec:appendixscatter}.

\subsection{Flux calibration}\label{appendixa1}
 Flux calibration of the pulsar is performed using the on source scan of a standard flux calibrator source 3C 286 and off source scan at 5$^\circ$ away from the flux calibrator to convert the data into physical flux density unit Jansky. We have used the following relation to calculate the sensitivity of the phased array for each 1 $\mathrm{MHz}$ frequency chunk for the 200 $\mathrm{MHz}$ band:
 \begin{equation}\label{eq:1}
\frac{P_{on}-P_{off}}{P_{off}}=S_{cal}\frac{G}{T_{sys}}  
\end{equation}
where $P_{on}$ is the mean power for the scan on 3C 286 and $P_{off}$ is the mean power while pointing the antennas at the cold sky 5$^\circ$ off from 3C 286, $T_{sys}$ is the overall system temperature in Kelvin, $S_{cal}$ is the flux density of calibrator source in jansky and $G$ is the antenna gain in Kelvins per jansky. We then multiply each 1 $\mathrm{MHz}$ frequency bin of baseline subtracted incoherently dedispersed and folded pulsar dataset with $\frac{T_{sys}}{G}$ estimated using Equation \ref{eq:1} to obtain a flux calibrated dataset.

\subsection{Calculating the Mean Flux Density} \label{appendixa2}
The mean flux density is obtained as the area under the pulse profile divided by the number of pulse phase bin ($n_{bin}$) in the pulse profile. We have accounted for the variable pulse width across frequency while calculating the mean flux density. The total error on the mean flux density is calculated as a quadratic sum of the off pulse rms noise ($\sigma_{off}$) scaled to the number of pulse phase bins in the profile ($\sigma~=~\frac{\sigma_{off}}{\sqrt{n_{bin}}}$) and a 10\% ($f_{err}$) flux scale uncertainty \citep{Chandra_2017}. Thus, total error on the mean flux density is 
\begin{equation}\label{eq:2}
\sigma_{total}=\sqrt{\sigma^2+(f_{err}\times S_{psr})^2} 
\end{equation}
where $S_{psr}$ is the mean flux density of the pulsar.

\subsection{Pulse Broadening and Scintillation}\label{subsec:appendixscatter}
The intrinsic pulse width of the pulsar radio emission gets broadened due to the dispersion and scattering. The effective pulse width is the quadratic sum of the intrinsic pulse width ($t_{intrinsic}$), pulse broadening due to dispersion measure ($t_{DM}$) and scattering ($t_s$). The effective pulse width can also be expressed in terms of the observed pulse profile as,
\begin{equation}\label{eq:3}
    t_{eff}=\sqrt{t_{intrinsic}^2+t_{DM}^2+t_s^2}=\frac{S_{mean}}{S_{peak}}P
\end{equation}
where $S_{mean}$ is the mean flux density, $S_{peak}$ is the peak flux density, and $P$ is the pulsar rotation period. We have estimated the effective pulse width at 345 $\mathrm{MHz}$ is 345 $\mathrm{\mu s}$. To obtain the scattering time scale we have assumed that the intrinsic pulse width is 10\% of the pulse period, $t_{intrinsic}=215.9$ $\mathrm{\mu s}$. 
Pulse broadening due intra-channel dispersion for finite channel width of 48.28 $\mathrm{kHz}$ is $t_{DM}=228$ $\mathrm{\mu s}$. Thus, using Equation \ref{eq:3}, the estimated value of the scattering timescale is $t_s\approx$ 141 $\mathrm{\mu s}$ and corresponding decorrelation bandwidth is $\Delta f=\frac{1}{2\pi t_s}\sim1.1$ $\mathrm{kHz}$.

\software{{\sc{gptool}} (Chowdhury et al., in preparation)\\
{\sc{dspsr}}\footnote{See http://dspsr.sourceforge.net/} \citep{dspsr2010}\\ {\sc{Tempo2}}\footnote{See https://bitbucket.org/psrsoft/tempo2/src/master/} \citep{Hobbs2006,Edwards2006}\\
{\sc{psrchive}}\footnote{See http://http://psrchive.sourceforge.net} \citep{Hotan2004}\\
{\it{scipy}}{\footnote{https://docs.scipy.org/doc/scipy/reference/generated/\\scipy.optimize.curve\_fit.html}} \citep{2020SciPy-NMeth}}

\facilities{GMRT\citep{Gupta_2017}}

\begin{acknowledgments}
We acknowledge support of the Department of Atomic Energy, Government of India, under project no.12-R\&D-TFR-5.02-0700. The GMRT is run by the National Centre for Radio Astrophysics of the Tata Institute of Fundamental Research, India. We thank the staff of the GMRT who have made these observations possible. We thank the anonymous reviewer for insightful comments and suggestions. D.K. acknowledge discussion with Aditya Chowdhury (NCRA-TIFR).
\end{acknowledgments}

\end{document}